\documentclass[12pt]{article}
\usepackage{amssymb}
\begin{document}
\begin{center}
{\Large \bf Algorithm for generating quasiperiodic packings\\ of 
icosahedral three-shell clusters}\\[5mm]
{\it Nicolae Cotfas}\\[5mm]
Faculty of Physics, University of Bucharest, Romania\\
E-mail: ncotfas@yahoo.com\\[1cm]
\end{center}
{\large \bf Introduction} \\[5mm]

The strip projection method is the most important way
to generate quasiperiodic patterns with predefined local structure.
We have obtained a very efficient algorithm for this method which 
allows one to use it in superspaces of very high dimension. 
A version of this algorithm for two-dimensional clusters and an 
application to decagonal two-shell clusters (strip projection in a
10-dimensional superspace) has been presented in  math-ph/0504036.  
The program in FORTRAN 90 used in this case is very fast
( 700-800 points are obtained in 3 minutes ).  

We present an application of our algorithm to three-dimensional
clusters. The physical three-dimensional space is embedded into a
31-dimensional superspace and the strip projection method is used in
order to generate a quasiperiodic packing of interpenetrating 
translated copies of a three-shell icosahedral cluster formed by
the 12 vertices of a regular icosahedron (the first shell), the
20 vertices of a regular dodecahedron (the second shell) and the
30 vertices of an icosidodecahedron (the third shell).

On a personal computer Pentium 4  with Fortran PowerStation version 4.0
(Microsoft Developer Studio) we obtain 400-500 points in 10 minutes.\\

More details, bibliography and samples can be found on the website:
\begin{center}
 http://fpcm5.fizica.unibuc.ro/~ncotfas/
\end{center}
\mbox{}\\[1cm]
{\large \bf Computer program in FORTRAN 90 and MATHEMATICA}\\
\begin{verbatim}
! QUASIPERIODIC PACKINGS OF THREE-SHELL ICOSAHEDRAL CLUSTERS
! *****(ICOSAHEDRON + DODECAHEDRON + ICOSIDODECAHEDRON)*****

! PLEASE INDICATE THE NUMBER OF POINTS YOU WANT TO ANALYSE
      INTEGER, PARAMETER :: N = 10000

! PLEASE INDICATE THE DIMENSION  M OF THE SUPERSPACE
      INTEGER, PARAMETER :: M = 31

      INTEGER  I, J, K, L, I1, I2, I3, I4, JJ, JP, JPP 
      REAL T, R1, R2, R3, D1, D2, D3, D4, AA
	  REAL, DIMENSION(M) :: V, W, TR
	  REAL, DIMENSION(3,3) :: C5
	  REAL, DIMENSION(3,M) :: B
	  REAL, DIMENSION(1:M-3,2:M-2,3:M-1,4:M) :: S
	  REAL, DIMENSION(N,M) :: P
	  REAL, DIMENSION(N) :: X, Y, Z

! PLEASE INDICATE THE RADIUS OF THE FIRST SHELL (ICOSAHEDRON)
	  R1 = 1.0

! PLEASE INDICATE THE RADIUS OF THE SECOND SHELL (DODECAHEDRON)
	  R2 = 1.2

 ! PLEASE INDICATE THE RADIUS OF THE THIRD SHELL (ICOSIDODECAHEDRON)
	  R3 = 1.5

! PLEASE INDICATE THE TRANSLATION OF THE STRIP YOU WANT TO USE	  
	  TR = 0.1

	  T = (1+SQRT(5.0))/2.0
	  C5(1,1) = (T-1)/2.0
	  C5(1,2) = -T/2.0
	  C5(1,3) = 1/2.0
	  C5(2,1) = T/2.0
	  C5(2,2) = 1/2.0
	  C5(2,3) = (T-1)/2.0
	  C5(3,1) = -1/2.0
	  C5(3,2) = (T-1)/2.0
	  C5(3,3) = T/2.0
	  B = 0.0
	  B(1,1) = R1 / SQRT(T+2.0)
	  B(2,1) = T * R1 / SQRT(T+2.0)
	  DO I = 2, 5 
	    DO J = 1, 3
		B(J,I) = SUM( C5(J,:) * B(:,I-1))
		END DO
	  END DO
	  B(2,6) = R1 / SQRT(T+2.0)
	  B(3,6) = T * R1 / SQRT(T+2.0)
      B(1,7) = R2 / SQRT(3.0)
	  B(2,7) = R2 / SQRT(3.0)
	  B(3,7) = R2 / SQRT(3.0)
	  DO I = 8, 11 
	    DO J = 1, 3
		B(J,I) = SUM( C5(J,:) * B(:,I-1))
		END DO
	  END DO
	  B(1,12) = R2 / SQRT(3.0)
	  B(2,12) = -R2 / SQRT(3.0)
	  B(3,12) = R2 / SQRT(3.0)
	  DO I = 13, 16 
	    DO J = 1, 3
		B(J,I) = SUM( C5(J,:) * B(:,I-1))
		END DO
	  END DO
	  B(1,17) = R3
	  DO I = 18, 21 
	    DO J = 1, 3
		B(J,I) = SUM( C5(J,:) * B(:,I-1))
		END DO
	  END DO
	  B(2,22) = R3
	  DO I = 23, 26 
	    DO J = 1, 3
		B(J,I) = SUM( C5(J,:) * B(:,I-1))
		END DO
	  END DO
	  B(3,27) = R3
	  DO I = 28, 31 
	    DO J = 1, 3
		B(J,I) = SUM( C5(J,:) * B(:,I-1))
		END DO
	  END DO
	  PRINT*, 'RADIUS OF THE FIRST SHELL (ICOSAHEDRON) IS ', R1
	  PRINT*, 'RADIUS OF THE SECOND SHELL (DODECAHEDRON) IS ', R2
	  PRINT*, 'RADIUS OF THE THIRD SHELL (ICOSIDODECAHEDRON) IS ', R3
	  PRINT*, 'STRIP TRANSLATED IN SUPERSPACE WITH THE VECTOR OF COORDINATES:'
	  PRINT*, TR
	  PRINT*, 'COORDINATES OF THE POINTS OF THREE-SHELL ICOSAHEDRAL CLUSTER'
	  PRINT*, '  (UP TO A SYMMETRY WITH RESPECT TO THE ORIGIN):'
	  DO J = 1, M
	  PRINT*, J, B(1,J), B(2,J), B(3,J)
	  END DO
	  PRINT*, 'PLEASE WAIT A FEW MINUTES OR MORE,& 
	           DEPENDING ON THE NUMBER OF ANALYSED POINTS'

      S = 0
      DO I1 = 1, M-3
      DO I2 = I1+1, M-2
      DO I3 = I2+1, M-1
      DO I4 = I3+1, M
	    DO D1 = -0.5, 0.5
		DO D2 = -0.5, 0.5		 
		DO D3 = -0.5, 0.5
		DO D4 = -0.5, 0.5
	    AA = D1 * ( B(1,I2) * B(2,I3) * B(3,I4) + &
		            B(2,I2) * B(3,I3) * B(1,I4) + &
					B(3,I2) * B(1,I3) * B(2,I4) - &
					B(3,I2) * B(2,I3) * B(1,I4) - &
		            B(1,I2) * B(3,I3) * B(2,I4) - &
					B(2,I2) * B(1,I3) * B(3,I4) ) - &
			 D2 * ( B(1,I1) * B(2,I3) * B(3,I4) + &
		            B(2,I1) * B(3,I3) * B(1,I4) + &
					B(3,I1) * B(1,I3) * B(2,I4) - &
					B(3,I1) * B(2,I3) * B(1,I4) - &
		            B(1,I1) * B(3,I3) * B(2,I4) - &
					B(2,I1) * B(1,I3) * B(3,I4) ) + &
			 D3 * ( B(1,I1) * B(2,I2) * B(3,I4) + &
		            B(2,I1) * B(3,I2) * B(1,I4) + &
					B(3,I1) * B(1,I2) * B(2,I4) - &
					B(3,I1) * B(2,I2) * B(1,I4) - &
		            B(1,I1) * B(3,I2) * B(2,I4) - &
					B(2,I1) * B(1,I2) * B(3,I4) ) - &
			 D4 * ( B(1,I1) * B(2,I2) * B(3,I3) + &
		            B(2,I1) * B(3,I2) * B(1,I3) + &
					B(3,I1) * B(1,I2) * B(2,I3) - &
					B(3,I1) * B(2,I2) * B(1,I3) - &
		            B(1,I1) * B(3,I2) * B(2,I3) - &
					B(2,I1) * B(1,I2) * B(3,I3) ) 
		IF ( AA > S(I1,I2,I3,I4) )  S(I1,I2,I3,I4) = AA
	  	END DO
        END DO
        END DO
        END DO
   IF( S(I1,I2,I3,I4) == 0) S(I1,I2,I3,I4) = N * SUM( B(1,:) ** 2 )
	  END DO
      END DO
      END DO
      END DO
      P = 0
	  P( 1,:) = ANINT( TR )
	  K = 1
	  L = 0
	  JP = 0
	  DO I = 1, N
	  IF( I <= K ) THEN 
	  V = P(I, : )	- TR
	  JJ = 1
	  JPP = 0
	    DO I1 = 1, M-3
        DO I2 = I1+1, M-2
        DO I3 = I2+1, M-1
        DO I4 = I3+1, M
	    AA = V(I1) * ( B(1,I2) * B(2,I3) * B(3,I4) + &
		               B(2,I2) * B(3,I3) * B(1,I4) + &
				       B(3,I2) * B(1,I3) * B(2,I4) - &
					   B(3,I2) * B(2,I3) * B(1,I4) - &
		               B(1,I2) * B(3,I3) * B(2,I4) - &
					   B(2,I2) * B(1,I3) * B(3,I4) ) - &
			 V(I2) * ( B(1,I1) * B(2,I3) * B(3,I4) + &
		               B(2,I1) * B(3,I3) * B(1,I4) + &
					   B(3,I1) * B(1,I3) * B(2,I4) - &
					   B(3,I1) * B(2,I3) * B(1,I4) - &
		               B(1,I1) * B(3,I3) * B(2,I4) - &
					   B(2,I1) * B(1,I3) * B(3,I4) ) + &
			 V(I3) * ( B(1,I1) * B(2,I2) * B(3,I4) + &
		               B(2,I1) * B(3,I2) * B(1,I4) + &
					   B(3,I1) * B(1,I2) * B(2,I4) - &
					   B(3,I1) * B(2,I2) * B(1,I4) - &
		               B(1,I1) * B(3,I2) * B(2,I4) - &
					   B(2,I1) * B(1,I2) * B(3,I4) ) - &
			 V(I4) * ( B(1,I1) * B(2,I2) * B(3,I3) + &
		               B(2,I1) * B(3,I2) * B(1,I3) + &
					   B(3,I1) * B(1,I2) * B(2,I3) - &
					   B(3,I1) * B(2,I2) * B(1,I3) - &
		               B(1,I1) * B(3,I2) * B(2,I3) - &
					   B(2,I1) * B(1,I2) * B(3,I3) ) 
		IF ( AA < -S(I1,I2,I3,I4) .OR. AA > S(I1,I2,I3,I4) )  JJ = 0
		IF ( AA == -S(I1,I2,I3,I4) .OR. AA == S(I1,I2,I3,I4) )  JPP = 1
		END DO
        END DO
        END DO
        END DO
	
		IF ( JJ .EQ. 1 ) THEN
		XP = 	SUM( V * B(1,:) )	
	    YP = 	SUM( V * B(2,:) )
		ZP = 	SUM( V * B(3,:) )
	    I3 = 1
		IF ( L > 0) THEN
	       DO J = 1, L	
	       IF( XP == X(J) .AND. YP == Y(J) .AND. ZP == Z(J) ) I3 = 0
	       END DO
		ELSE
		END IF
	    IF( I3 == 1) THEN	
	    IF( JPP .EQ. 1 ) JP = JP + 1	 							  
        L = L + 1
		X(L) = XP	  
        Y(L) = YP
		Z(L) = ZP 
		ELSE
	    END IF 
		DO I1 = 1, M
		 DO I2 = -1, 1
		  W = P(I,:)
		  W(I1) = W(I1) + I2
		  I3 = 0
		   DO J = 1, K
		   IF( ALL(W .EQ. P(J,:)) ) I3 = 1
           END DO
		    IF( I3 == 0 .AND. K < N ) THEN
			K = K + 1
			P(K, : ) = W
			ELSE
			END IF
		 END DO
		END DO
		ELSE
	   END IF
	  ELSE
	  END IF
	  END DO
	  PRINT*, 'NUMBER OF ANALYSED POINTS : ', K
	  PRINT*, 'NUMBER OF OBTAINED POINTS : ', L
      PRINT*, 'NUMBER OF POINTS LYING ON THE FRONTIERE OF THE & 
                                                  STRIP:', JP
	  PRINT*, 'PLEASE INDICATE THE NAME OF A FILE FOR RESULTS'	  
      WRITE(4,98)					  
  98  FORMAT('Show[Graphics3D[{ PointSize[0.01],{') 
	  DO J = 1, L-1
	  WRITE(4,99) X(J), Y(J), Z(J)		
  99  FORMAT( 'Point[{'F10.5','F10.5','F10.5,'}], ')
	  END DO
  	  WRITE(4,100) X(L), Y(L), Z(L)
 100  FORMAT( 'Point[{'F10.5','F10.5','F10.5,'}]')
	  WRITE(4,101)
 101  FORMAT('}} ]]')
	PRINT*, '* OPEN THE FILE CONTAINING THE RESULTS WITH &
                                                "NotePad" '
	PRINT*, '* SELECT THE CONTENT OF THE FILE ("Select All") &
                                         AND COPY IT ("Copy")'
	PRINT*, '* OPEN "MATHEMATICA", PASTE THE COPIED FILE, &
                               AND EXECUTE IT ("Shift+Enter")'

	  END
\end{verbatim}  
\end{document}